RUBBISH DNA: THE FUNCTIONLESS FRACTION OF THE HUMAN GENOME

Dan Graur

University of Houston

*Abstract*

Because genomes are products of natural processes rather than "intelligent design," all genomes contain functional and nonfunctional parts. The fraction of the genome that has no biological function is called "rubbish DNA." Rubbish DNA consists of "junk DNA," i.e., the fraction of the genome on which selection does not operate, and "garbage DNA," i.e., sequences that lower the fitness of the organism, but exist in the genome because purifying selection is neither omnipotent nor instantaneous. In this chapter, I (1) review the concepts of genomic function and functionlessness from an evolutionary perspective, (2) present a precise nomenclature of genomic function, (3) discuss the evidence for the existence of vast quantities of junk DNA within the human genome, (4) discuss the mutational mechanisms responsible for generating junk DNA, (5) spell out the necessary evolutionary conditions for maintaining junk DNA, (6) outline various methodologies for estimating the functional fraction within the genome, and (7) present a recent estimate for the functional fraction of our genome.

*Introduction*

While evolutionary biologists and population geneticists have been comfortable with the concept of genomic functionlessness for more than half a century, classical geneticists and their descendants have continued to exist in an imaginary engineered world, in which each and every nucleotide in the genome is assumed to have a function and evolution counts for naught. Under this pre-Darwinian mindset, for instance, Vogel (1964) estimated the human genome to contain approximately 6.7 million protein-coding genes. Interestingly, Vogel (1964) deemed this number to be "disturbingly high," yet he could not bring himself to admit even a small fraction of "meaningless" DNA. Instead, he postulated one of two possibilities: either the protein-coding genes in humans are 100 times larger than those in bacteria, or "systems of higher order which are connected with structural genes in operons and regulate their activity" occupy a much larger part of the genetic material than do protein-coding genes.

Since 1964, the number of protein-coding genes in the human genome has come down considerably, in a process that has been at times extremely humbling. Moreover, scientists discovered that many other species, including plants and unicellular organisms, have more genes than we do (Pertea and Salzberg 2010). Some estimates for the number of protein-coding genes before the Human Genome Project ranged from 100,000 to more than half a million. These estimates were drastically reduced with the publication in 2001 of the draft human genome, in which the number was said to be 26,000-30,000 (Lander et al. 2001). In 2004, with the publication of the "finished" euchromatic sequence of the human genome (International Human Genome Sequencing Consortium 2004), the number was reduced to ~24,500, and in 2007 it further decreased to ~20,500 (Clamp et al. 2007). The lowest ever estimate was 18,877 (Pruitt et al. 2009), which agrees quite well with the newest estimate for the number of protein-coding genes in the human nuclear genome (Ezkurdia et al. 2014).

Protein-coding genes turned out to occupy approximately 2% of the human genome. Of course, no one in his right mind thought that all the 98% or so of the human genome that does not encode proteins is functionless. Revisionist claims that equate noncoding DNA with junk DNA (e.g., Hayden 2010) merely reveal that people who are allowed to exhibit their logorrhea in *Nature* and other glam journals are as ignorant as the worst young-earth creationists. The question of genome functionality can be phrased qualitatively (Does the human genome contain functionless parts?) or quantitatively (What proportion of the human genome is functional?). Among people unversed in evolutionary biology (e.g., ENCODE Project Consortium 2012), a misconception exists according to which genomes that are wholly functional can be produced by natural processes. Actually, for the evolutionary process to produce a wholly functional genome, several conditions must be met: (1) the effective population size needs to be enormous—infinite to be precise, (2) the deleterious effects of increasing genome size by even a single nucleotide should be considerable, and (3) the generation time has to be very short. In other words, never! Not even in the commonest of bacterial species population on Earth are these conditions met, let alone in species with small effective population sizes and long generation times such as perennial plants and humans. A genome that is 100% functional is a logical impossibility.

*What is Function?*

Like many words in the English language, "function" has numerous meanings. In biology, there are two main concepts of function: the "selected effect" and "causal role." The selected-effect function, also referred to as the proper-biological function, is a historical concept. In other words, it explains the origin, the cause (etiology), and the subsequent evolution of the trait (Millikan 1989; Neander 1991). Accordingly, for a trait, *T*, to have a selected-effect function, *F*, it is necessary and sufficient that the following two conditions hold: (1) *T* originated as a "reproduction" (a copy or a copy of a copy) of some prior trait that performed *F* (or some function similar to *F*, say *F'*) in the past, and (2) *T* exists because of *F* (Millikan 1989). In other words, the selected-effect function of a trait is the effect for which the trait was selected and by which it is maintained. The selected-effect function answers the question: Why does *T* exist?

The causal-role function is ahistorical and nonevolutionary (Cummins 1975; Amundson and Lauder 1994). That is, for a trait, *Q*, to have a causal-role function, *G*, it is necessary and sufficient that *Q* performs *G*. The causal-role function answers the question: What does *Q* do? Most biologists follow Dobzhansky's dictum according to which biological meaning can only be derived from evolutionary context. Hence, with few exceptions, they use the selected-effect concept of function. We note, however, that the causal-role concept may sometimes be useful, for example, as an *ad hoc* device for traits whose evolutionary history and underlying biology are obscure. Furthermore, we note that all selected-effect functions have a causal role, while the vast majority of causal-role functions do not have a selected-effect function. It is, thus, wrong to assume that all causal-role functions are biologically relevant. Doolittle et al. (2014), for instance, prefers to restrict the term "function" to selected-effect function, and to refer to causal-role function as "activity."

Using the causal-role concept of function in the biological sciences can lead to bizarre outcomes. For example, while the selected-effect function of the heart can be stated unambiguously to be the pumping of blood, the heart may be assigned many additional causal-role functions, such as adding 300 grams to body weight, producing sounds, preventing the pericardium from deflating onto itself, and providing an inspiration for love songs and Hallmark cards (Graur et al. 2013). The thumping noise made by the heart is a favorite of philosophers of science; it is a valuable aid in medical diagnosis, but it is not the evolutionary reason we have a heart. An even greater absurdity of using the causal-role concept of function arises when realizing that every nucleotide in a genome has a causal role—it is replicated! Does that mean that every nucleotide in the genome has evolved for the purpose of being copied?

Distinguishing between what a genomic element does (its causal-role activity) from why it exists (its selected-effect function) is a very important distinction in biology (Huneman 2013; Brunet and Doolittle 2014). Ignoring this distinction, and assuming that all genomic sites that exhibit a certain biochemical activity are functional, as was done, for instance, by the ENCODE Project Consortium (2012), is equivalent to claiming that following a collision between a car and a pedestrian, a car's hood would be ascribed the "function" of harming the pedestrian while the pedestrian would have the "function" of denting the car's hood (Hurst 2012).

The main advantage of the selected-effect function definition is that it suggests a clear and conservative method for inferring function in a DNA sequence—only sequences that can be shown to be under selection can be claimed with any degree of confidence to be functional. From an evolutionary viewpoint, a function can be assigned to a DNA sequence if and only if it is possible to destroy it (Graur et al. 2013). All functional entities in the universe can be rendered nonfunctional by the ravages of time, entropy, mutation, and what have you. Unless a genomic functionality is actively protected by selection, it will accumulate deleterious mutations and will cease to be functional. The absurd alternative is to assume that function can be assessed independently of selection, i.e., that no deleterious mutation can ever occur in the region that is deemed to be functional. Such an assumption is akin to claiming that a television set left on and unattended will still be in working condition after a million years because no natural events, such as rust, erosion, static electricity, and the gnawing activity of rodents can affect it (Graur et al. 2103). A convoluted "rationale" for discarding natural selection as the arbiter of functionality was put forward by Stamatoyannopoulos (2012). This paper should be read as a cautionary tale of how genome biology has been corrupted by medical doctors and other ignoramuses who uncritically use the causal-role concept of function.

Function should always be defined in the present tense. In the absence of prophetic powers, one cannot use the potential for creating a new function as the basis for claiming that a certain genomic element is functional. For example, the fact a handful of transposable elements have been coopted into function cannot be taken as support for the hypothesis that all transposable elements are functional. In this respect, the Aristotelian difference between "potentiality" and "actuality" is crucial.

What is the proper manner in which null hypotheses concerning the functionality or nonfunctionality of a particular genomic element should be phrased? Most science practitioners adhere to Popper's system of demarcation according to which scientific progress is achieved through the falsification of hypotheses that do not withstand logical or empirical tests. Thus, a null hypothesis should be phrased in such a manner as to spell out the conditions for its own refutation. Should one assume lack of functionality as the null hypothesis, or should one assume functionality? Let us consider both cases. A statement to the effect that a genomic element is devoid of a selected-effect function can be easily rejected by showing that the element evolves in a manner that is inconsistent with the expectations of strict neutrality. If, on the other hand, one assumes as the null hypothesis that an element is functional, then failing to find telltale indicators of selection cannot be interpreted as a rejection of the hypothesis, merely as a sign that we have not searched thoroughly enough or that the telltale sign of selection have been erased by subsequent evolutionary events.

There exists a fundamental asymmetry between verifiability and falsifiability in science: scientific hypotheses can never be proven right; they can only be proven wrong. The hypothesis that a certain genomic element is functional can never be rejected and is, hence, unscientific. According to physicist Wolfgang Pauli (quoted in Peierls 1960) a hypothesis that cannot be refuted "is not only not right, it is not even wrong."

*An Evolutionary Classification of Genome Activity and Genome Function*

In terms of its biochemical activities, a genome can be divided into three classes: (1) regions that are transcribed and translated, (2) regions that are transcribed but not translated, and (3) regions that are not transcribed (Figure 1). Each of these three classes can be either functional or functionless. Activity has nothing to do with function. A genome segment may be biochemically active yet have no biologically meaningful function. An analogous situation arises when my shoe binds a piece of chewing gum during hot days in Houston. The binding of the chewing gum is an observable activity, but no reasonable person would claim that this is the function of shoes.

Genomic sequences are frequently referred to by their biochemical activity, regardless of whether or not such activity is biologically meaningful. The need for a rigorous evolutionary classification of genomic elements by selected-effect function arises from two erroneous and sometimes deliberately disingenuously equivalencies that are frequently found in the literature. The first equivalency, usually espoused in the medical and popular literature (e.g., Krams and Bromberg 2013; Mehta et al. 2013), misleadingly and inappropriately synonymizes "noncoding DNA"—i.e., all regions in the genome that do not encode proteins—with "junk DNA"—i.e., all regions in the genome that are neither functional nor deleterious. The second, even more pernicious equivalency transmutes every biochemical activity into a function (e.g., ENCODE Project Consortium 2012; Sundaram et al. 2014; Kellis et al. 2014).

The classification scheme presented below is based on Graur et al. (2015). The genome is divided into "functional DNA" and "rubbish DNA" (Figure 2). Functional DNA refers to any segment in the genome whose selected-effect function is that for which it was selected and/or by which it is maintained. Most functional sequences in the genome are maintained by purifying selection. Less frequently, functional sequences exhibit telltale signs of either positive or balancing selection. A causal-role activity, such as "low-level noncoding-RNA transcription" (e.g., Kellis et al. 2014), is an insufficient attribute of functionality.

Functional DNA is further divided into "literal DNA" and "indifferent DNA." In literal DNA, the order of nucleotides is under selection. Strictly, a DNA element of length $l$ is defined as literal DNA if its function can be performed by a very small subset of the $4^l$ possible sequences of length $l$. For example, there are three possible sequences of length 3 that can encode isoleucine according to the standard genetic code, as opposed to the much larger number, 64, of possible three-nucleotide sequences. Functional protein-coding genes, RNA-specifying genes, and untranscribed control elements are included within this category.

Indifferent DNA includes genomic segments that are functional and needed, but the order of nucleotides in their sequences is of little consequence. In other words, indifferent DNA refers to sequences whose main function is being there, but whose exact sequence is not important. They serve as spacers, fillers, protectors against frameshifts, and may possess nucleotypic functions, such as determining nucleus size. The third codon position in four-fold degenerate codons may be regarded as a simple example of indifferent DNA; the nucleotide that resides at this position is unimportant, but the position itself needs to be occupied. Thus, indifferent DNA should show no evidence of selection for or against point mutations, but deletions and insertions should be under selection.

Rubbish DNA (Brenner 1998) refers to genomic segments that have no selected-effect function. Rubbish DNA can be further subdivided into "junk DNA" and "garbage DNA." We have written evidence that the term "junk DNA" was already in use in the early 1960s (e.g., Aronson et al. 1960; Ehret and de Haller 1963), however, it was Susumu Ohno (1972, 1973) who formalized its meaning and provided an evolutionary rationale for its existence. "Junk DNA" refers to a genomic segment on which selection does not operate and, hence, it evolves neutrally. Of course, some junk DNA may acquire a useful function in the future, although such an even is expected to occur only very rarely. Thus, the "junk" in "junk DNA" is identical in its meaning to the colloquial "junk," such as when a person mentions a "garage full of junk," in which the implications are: (1) that the garage fulfills its intended purpose, (2) that the garage contains useless objects, and (3) that in the future some of the useless objects may (or may not) become useful in the future. Of course, as in the case of the garage full of junk, the majority of junk DNA will never acquire a function. Junk DNA and the junk in one's garage are also similar in that "they may be kept for years and years and, then, thrown out a day before becoming useful" (David Wool, personal communication).

The term "junk DNA" has generated a lot of controversy. First, because of linguistic prudery and the fact that "junk" is used euphemistically in off-color contexts, some biologists find the term "junk DNA" "derogatory" and "disrespectful" (e.g., Brosius and Gould 1992). An additional opposition to the term "junk DNA" stems from false teleological reasoning. Many researchers (e.g., Makalowski 2003; Wen et al. 2012) use the term "junk DNA" to denote a piece of DNA that can never, under any evolutionary circumstance, be selected for or against. Since every piece of DNA may become advantageous or deleterious by gain-of-function mutations, this type of reasoning is indefensible. A piece of junk DNA may indeed be coopted into function, but that does not mean that it will be, let alone that it currently has a function. Finally, some opposition to the term is related to the anti-scientific practice of assuming functionality as the null hypothesis (Petsko 2003).

Garbage DNA refers to sequences that exist in the genome despite being actively selected against. The reason that detrimental sequences are observable is that selection is neither omnipotent nor rapid. At any slice of evolutionary time, segments of garbage DNA (presumably on their way to becoming extinct) may be found in the genome. The distinction between junk DNA and garbage DNA was suggested by Brenner (1998):

> *"Some years ago I noticed that there are two kinds of rubbish in the world and that most languages have different words to distinguish them. There is the rubbish we keep, which is junk, and the rubbish we throw away, which is garbage. The excess DNA in our genomes is junk, and it is there because it is harmless, as well as being useless, and because the molecular processes generating extra DNA outpace those getting rid of it. Were the extra DNA to become disadvantageous, it would become subject to selection, just as junk that takes up too much space, or is beginning to smell, is instantly converted to garbage by one's wife, that excellent Darwinian instrument."*

Each of the four functional categories described above can be (1) transcribed and translated, (2) transcribed but not translated, or (3) not transcribed. Hence, we may encounter, for instance, junk DNA, literal RNA, and garbage proteins.

*Changes in Functional Affiliation*

The affiliation of a DNA segment to a particular functional category may change during evolution. With four functional categories, there are twelve possible such changes. Several such changes are known to occur quite frequently (Figure 3). For example, junk DNA may become garbage DNA if the effective population size increase; the opposite will occur if the effective population size decreases (Ohta 1973). Many of the twelve possible changes have been documented in the literature. Pseudogenes, for instance, represent a change in functional status from literal DNA to junk DNA, while some diseases are caused by either a change from functional DNA to garbage DNA (e.g., Chen et al. 2003) or from junk DNA to garbage DNA (Cho and Brant 2011).

Rubbish DNA mutating to functional DNA may be referred to as "Lazarus DNA," so named after Lazarus of Bethany, the second most famous resurrected corpse in fiction (John 11:38-44; 12:1; 12:9; 12:17). Similarly, functional DNA may mutate to garbage DNA, in which case the term "Jekyll-to-Hyde DNA," based on the fictional transformation of a benevolent entity into a malicious one (Stevenson 1886), was suggested. Garbage DNA may also be derived from junk DNA, for which the term "zombie DNA" seems appropriate (Kolata 2010).

*Genome Size Variation*

Because "genome size" and "DNA content" are used inconsistently in the literature (Greilhuber et al. 2005), when referring to the size of the haploid genome it is advisable to use the unambiguous term "C value" (Swift 1950), where "C" stands for "constant," to denote the fact that the intraspecific variability in haploid genome size is substantially smaller than the interspecific variability.

With very few exceptions eukaryotes have much larger genomes than prokaryotes. C values in eukaryotes range from less than $2.3 \times 10^6$ bp in the microsporidian *Encephalitozoon intestinalis* (Corradi et al. 2010) to approximately $1.5 \times 10^{11}$ bp in the canopy plant *Paris japonica* (Pellicier et al. 2010). If one includes organelles of eukaryotic origin, genomes in eukaryotes span a close to a 400,000-fold size range.

What can explain the huge variation in eukaryotic genome size? Let us first investigate if any genomic compartment correlates with genome size. First, we note that in contradistinction to the situation in prokaryotes, only a miniscule fraction of the eukaryotic genome is occupied by protein-coding sequences. Moreover, the number of protein-coding genes does not correlate with genome size. To illustrate this fact let us compare the human genome with the genome of a teleost fish called *Takifugu rubripes* (formerly *Fugu rubripes*). At 400 Mb, the *T. rubripes* genome is one of the smallest vertebrate genomes (Aparicio et al. 2002; Noleto et al. 2009). Interestingly, although the length of the *Takifugu* genome is less than about one-eighth that of the human genome, it contains a comparable number of protein-coding genes (Aparicio et al. 2002). It is for this reason that Sydney Brenner regarded the *Takifugu* genome as "the *Reader's Digest* version" of the human genome (quoted in Purves et al. 2004). The small genome size of *Takifugu* can be attributed to a reduction in intron and intergenic lengths, a lack of significant amounts of repetitive sequences, and a very small number of pseudogenes—the genome of *Takifugu rubripes* has no mitochondrial pseudogenes and only 162 nuclear pseudognes versus at least 15,000 pseudogenes in the human genome (Hazkani-Covo et al. 2010). Many other small-genome organisms have *Takifugu*-like characteristics. For example, the carnivorous bladderwort plant *Utricularia gibba* has a tiny genome, yet it accommodates a typical number of protein-coding genes (~30,000) for a plant (Ibarra-Lachlette et al. 2013).

Can genome size variation be explained by other variables related to protein-coding genes? While there are small differences in the mean mRNA length among different organisms, no correlation exists between this variable and genome size. For instance, mRNAs are slightly longer in multicellular organisms than in protists (1,400–2,200 bp versus 1,200–1,500 bp), whereas genome sizes in protists can be much larger than those of multicellular organisms. Similarly, organisms with larger genomes do not always produce larger proteins, nor do they have longer introns.

Finally, while there exists a significant positive correlation between the degree of repetition of several RNA-specifying genes and genome size, these genes constitute only a negligible fraction of the eukaryotic genome, such that the variation in the number of RNA-specifying genes cannot explain the variation in genome size.

*The C-value Paradox as Evidence for the Existence of Junk DNA within the Human Genome*

A genomic paradox refers to the lack of correspondence between a measure of genome size or genome content and the presumed amount of genetic information "needed" by the organism, i.e., its complexity or organismal complexity. Different measures of complexity have been put forward in different fields. However, because researchers ask similar questions about the complexity of their different subjects of research, the answers that they came up with for how to measure complexity in biology, computer science, or finance bear a considerable similarity to one another (Lloyd 2001). Measures of complexity of an entity attempt to answer three questions: (1) How hard is it to describe? (2) How hard is it to create or evolve? and (3) What is its degree of organization?

The simplest measures of organismal complexity are the number of cells and the number of cell types (Kaufman 1971). The vast majority of eukaryotic taxa are unicellular, i.e., they have a single cell and a single cell type. In animals, the number of cells varies between $10^3$ in some nematodes to $10^{12}$ in some mammals; the number of cell types varies from 4 and 11 in placozoans and sponges, respectively, to approximately 200 in mammals (Klotzko 2001; Srivastava et al. 2008; Goldberg 2013). The number of cell types in plants is about one order of magnitude smaller than that in mammals (Burgess 1985). Thus, organismal complexity as measured by the number of cell types ranges from 1 to approximately 200 whereas C-values range over a 400,000-fold range. Numerous lines of evidence indicate that there is no relationship between genome size and organismal complexity. For example, many unicellular protozoans, which by definition have a single cell and a single cell type, possess much larger genomes than mammals, which possess trillions of cells and 200 or more cell types, and are presumably infinitely more "sophisticated" than protists (Table 1). The same goes for the comparison between plants and mammals. Many plants have much larger genomes than mammals while at the same time possessing fewer cell types. Furthermore, organisms that are similar in morphological and anatomical complexity (e.g., flies and locusts, onion and lily, *Paramecium aurelia* and *P. caudatum*) exhibit vastly different C values (Table 1). This lack of a positive correlation between organismal complexity and genome size is also evident in comparisons of sibling species (i.e., species that are so similar to each other morphologically as to be indistinguishable phenotypically). In insects, protists, bony fishes, amphibians, and flowering plants, many sibling species differ greatly in their C-values, even though by definition no difference in organismic complexity exists. The example of sibling species is extremely illuminating since it tells that the same level of complexity can be achieved by vastly different amounts of genomic DNA.

This lack of correspondence between C values and the presumed complexity of the organisms has become known in the literature as the C-value paradox (Thomas 1971) or the C-value enigma (Gregory 2001). There are two facets to the C value paradox. The first concerns the undisputable fact that genome size cannot be used as a predictor of organismal complexity. That is, from knowledge of the C value it is impossible to say whether an organism is unicellular or multicellular, whether the genome contains few or many genes, or whether the organism is made of a few or a couple of hundreds of cell types. The second facet reflects an unmistakably anthropocentric bias. For example, some salamanders have genome sizes that are almost 40 times bigger than that of humans. So, although there is no definition of organismal complexity that shows that salamanders are objectively less complex than humans (Brookfield 2000), it is still very difficult for us humans to accept the fact that our species, from the viewpoint of genome size, does not look at all like the "pinnacle of creation" and the "paragon of animals." Realizing that our genome is so much smaller than those of frogs, sturgeons, shrimp, squids, flatworms, mosses, onions, daffodils, and amoebas can be quite humbling, if not outright insulting. Moran (2007) referred to the anthropocentric difficulties of some people to accept their reduced genomic status as the "deflated-ego problem."

The C-value paradox requires us to explain why some very complex organisms have so much less DNA than less complex ones. Why do humans have less than half the DNA in the genome of a unicellular ciliate? Since it would be illogical to assume that an organism possesses less DNA than the amount required for its vital functions, the logical inference is that many organisms have vast excesses of DNA over their needs. Hence, large genomes must contain unneeded and presumably functionless DNA. This point of view, incidentally, is not very popular with people who regard organism as epitomes of perfection either because they believe in creationism directed by an omnipotent being or because they lack proper training in evolutionary biology and erroneously believe that natural selection is omnipotent.

As noted by Orgel and Crick (1980), if one assumes that genomes are 100% functional, then one must conclude that the number of genes needed by a salamander is 20 times larger than that in humans. It should be noted that organisms that have much larger genomes than humans are neither rare nor exceptional. For example, more than 200 salamander species have been analyzed thus far, and all their genomes have been found to be 4-35 times larger than the human genome.

These observations pose an insurmountable challenge to any claim that most eukaryotic DNA is functional. The challenge is beautifully illustrated by "the onion test" (Palazzo and Gregory 2014). The domestic onion, *Allium cepa*, is a diploid plant (2n = 16) with a haploid genome size of roughly 16 Gb, i.e., approximately five times larger than that of humans. (The example of the onion was chosen arbitrarily, presumably for its shock value—any other species with a large genome could have been chosen for this comparison.) The onion test simply asks: if most DNA is functional, then why does an onion require five times more DNA than a human?

If one assumes that the human genome is entirely functional, then the human genome becomes the Goldilocks of genomes—not too small for the organism who defines itself as the "pinnacle of creation," yet not too big for it to become littered with functionless DNA. The onion genome, on the other hand, would by necessity be assumed to possess junk DNA lest we regard onions as our superiors. The C-value paradox and the rejection of our genomic status as Goldilocks inevitably lead us to the conclusion that we possess junk DNA.

*Genetic Mutational Load: Can the Human Genome be 100% Functional?*

Many evolutionary processes can cause a population to have a mean fitness lower than its theoretical maximum. For example, deleterious mutations may occur faster than selection can get rid of them, recombination may break apart favorable combinations of alleles creating less fit combinations, and genetic drift may cause allele frequencies to change in a manner that is antagonistic to the effects of natural selection. Genetic load is defined as the reduction in the mean fitness of a population relative to a population composed entirely of individuals having the maximal fitness. The basic idea of the genetic load was first discussed by Haldane (1937) and later by Muller (1950).

If $w_{max}$ is the fitness of the fittest possible genotype and $\bar{w}$ is the mean fitness of the actual population, then the genetic load ($L$) is the proportional reduction in mean fitness relative to highest possible fitness:

$$L = \frac{w_{max} - \bar{w}}{w_{max}} \qquad (1)$$

Here, we shall consider the mutational genetic load, i.e., the reduction in mean population fitness due to deleterious mutations. Haldane (1937) showed that the decrease of fitness in a species as a consequence of recurrent mutation is approximately equal to the total mutation rate multiplied by a factor that ranges from 1 to 2 depending on dominance, inbreeding, and sex linkage. For example, for a recessive mutation, it has been shown that as long as the selective disadvantage of the mutant is larger than the mutation rate and the heterozygote fitness is neither larger nor smaller that the fitness values of the homozygotes, the mutational load is approximately equal to the mutation rate (Kimura 1961; Kimura and Maruyama 1966). Thus,

$$L = \mu \qquad (2)$$

and

$$\bar{w} = (1-\mu)^n \tag{3}$$

where *n* is the number of loci. In the following we only deal with recessive mutations. We note, however, that under the same conditions, the mutational load for a dominant mutation is approximately twice the mutation rate.

In randomly mating populations at equilibrium, the mutational load does not depend on the strength of the selection against the mutation. This surprising result comes from the fact that alleles under strong selection are relatively rare, but their effects on mean fitness are large, while the alleles under weak purifying selection are common, but their effects on mean fitness are small; the effects of these two types of mutation neatly cancel out. As a result, in order to understand the magnitude of mutation load in randomly mating populations, we conveniently need only know the deleterious mutation rate, not the distribution of fitness effects.

Let us now consider the connection between mutational genetic load and fertility. The mean fertility of a population ($\bar{F}$) is the mean number of offspring born per individual. If the mortality rate before reproduction age is 0 and mean fertility is 1, then the population will remain constant in size from generation to generation. In real populations, however, the mortality rate before reproduction is greater than 0 and, hence, mean fertility needs to be larger than 1 to maintain a constant population size. In the general case, for a population to maintain constant size its mean fertility should be

$$\bar{F} = \frac{1}{\bar{w}} \qquad (4)$$

Let us consider the following example from Nei (2013). Assume that there are 10,000 loci in the genome and that the mutation rate is $\mu = 10^{-5}$ per locus per generation. Under the assumption that all mutants are deleterious and recessive, the mutational load is $L = 10{,}000 \times 10^{-5} = 0.1$ and the mean fitness of the population is $\bar{w} = \left(1 - 10^{-5}\right)^{10{,}000} \approx 0.90$. Therefore, the average fertility is $\bar{F} = \frac{1}{0.90} = 1.11$. That is, each individual should have on average 1.11 descendants for the population to remain constant in size. For a mammal, like humans, such a load is easily bearable.

Let us now assume that the entire diploid human genome ($2 \times 3 \times 10^9$ bp) is functional, and contains functional elements only. If the length of each functional element is the same as that of a bacterial protein-coding gene, i.e., ~1,000 nucleotides, then the human genome should consist of approximately 3 million functional loci. With a mutation rate of $10^5$ per locus per generation, the total mutational load would be $L = 30$ and the mean population fitness is $9 \times 10^{-14}$. The average individual fertility required to maintain such a population will be $\bar{F} = 1.1 \times 10^{13}$. That is, each individual in the population would have to give birth to 11,000,000,000,000 children and all but one would die before reproductive age. This number is absurdly high. Muller (1950) suggested that genetic load values cannot exceed $L = 1$. As a matter of fact, he believed that the human genome has no more than 30,000 genes, i.e., a genetic load of $L = 0.3$, an average fitness of $\bar{w} = 0.72$, and an average fertility per individual of $\bar{F} = 1.39$. Indeed, modern estimates of mutational genetic load in human populations are lower or much lower than 1 (Keightley and Eyre-Walker 2000; Keightley 2012; Simons et al. 2014).

In the above, we assumed that deleterious mutations have an additive effect on fitness. Any factor that increases the number of deleterious mutations removed from the population, such as negative epistasis or inbreeding, will reduce the mutational load. On the other hand, any factor that decreases the efficacy of selection, such as positive epistasis or reduction in effective population size, will increase the mutational genetic load. Be that as it may, let us now consider the implications of the mutational genetic load on the fraction of the genome that is functional.

Studies have shown that the genome of each human newborn carries 56-103 point mutations that are not found in either of the two parental genomes (Xue et al. 2009; Roach et al. 2010; Conrad et al. 2011; Kong et al. 2012). If 80% of the genome is functional, as trumpeted by ENCODE Project Consortium (2012), then 45-82 deleterious mutations arise per generation. For the human population to maintain its current population size under these conditions, each of us should have on average $3 \times 10^{19}$ to $5 \times 10^{35}$ (30,000,000,000,000,000,000 to 500,000,000,000,000,000,000,000,000,000,000,000) children. This is clearly bonkers. If the human genome consists mostly of junk and indifferent DNA, i.e., if the vast majority of point mutations are neutral, this absurd situation would not arise.

*Detecting Functionality at the Genome Level*

The availability of intraspecific and interspecific genomic sequences has made it possible not only to test whether or not a certain genomic region is subject to selection, but also to exhaustively scan the genome for regions likely to have been the target of selection and are, hence, of functional importance. For a given species, such scans allow us to estimate the proportion of the genome that is functional. In the literature, numerous approaches for detecting selection through comparisons of DNA sequences have been proposed (e.g., Nielsen 2005; Andrés et al. 2009; Li 2011; Vitti et al. 2013; Grossman et al. 2013; Lawrie and Petrov 2014). The main difference between tests designed to detect selection at a particular locus and tests involving genome-wide comparisons is that the latter involve multiple tests at multiple loci. Thus, many of the sites that are identified through such methods as having been subjected to selection are expected to be false positives. The statistics must, hence, be adjusted to the number of tests performed, using either standard techniques for multiple comparisons or adjusting significance levels to account for false discovery rates (e.g., Massingham and Goldman 2005).

Some methods for detecting selection at the genomic level require comparisons among species, some rely on intraspecific comparisons, and yet others require both types of data. Some methods are applicable to protein-coding genes only; some are applicable to all sequences. Some are based on comparisons of allele frequencies, some are based on linkage disequilibrium measures, and some rely on population-differentiation measures, such as genetic distances. Some are suitable for detecting purifying selection, and some are suitable for detecting positive or balancing selection. A straightforward method of estimating the functional fraction of a genome is to add up the genomic fractions that are under (1) positive, (2) negative, and (3) balancing selection.

Detecting functional regions subject to purifying selection is relatively straightforward. In interspecific comparisons, homologous genomic regions under purifying selection are expected to be more similar to one another than unselected regions. The rationale for this expectation is that many mutations in functional sequences are deleterious and, hence, weeded out of the population. Thus, purifying selection is primarily observable as highly conserved regions. We note, however, that evolutionary processes other than selection, such as mutational coldspots and gene conversion, can result in sequence conservation (Ahituv et al. 2007).

Balancing selection at the level of the genome is one of the least studied areas in evolutionary genomics. This lack of interest is somewhat unexpected given that a strong association between balancing selection and pathology has been hypothesized for a large number of human diseases, such as sickle-cell anemia, cystic fibrosis, and phenylketonuria. Andrés et al. (2009) devised a method for identifying regions that concomitantly show excessive levels of nucleotide polymorphism (as measured by the number of polymorphic sites in a gene), as well as an excess of alleles at intermediate allele frequencies.

All in all, the main emphasis in the last two decades has been with methods intended to detect positive selection. The principal reason for the emphasis on positive selection at the expense of purifying and balancing selection is that positive Darwinian selection is presupposed to be the primary mechanism of adaptation. Detecting genomic regions that have experienced positive selection requires more nuanced procedures and significantly more data than those required by methods for detecting purifying selection. Moreover, the various methods for detecting positive selection at the level of the genome are known to yield many false positives. Although each method has its own particular strengths and limitations, there are a number of challenges that are shared among all tests. First, deviations from neutrality expectations may be explained by factors other than selection. Demographic events, such as migration, expansions, and bottlenecks, can often yield signals that mimic selection. This recognition has led some researchers to adopt approaches that explicitly attempt to separate demographic effects from selection effects (Li and Stephan 2006; Excoffier et al. 2009). Second, even when confounding effects are dealt with, the interpretation of selection may not be straightforward. For example, rate-based tests identify as "functional" all regions in which evolutionary rates have been accelerated. Such regions may indeed be subject to positive selection, but the acceleration may also be due to (1) the relaxation of selective constraint following total or partial nonfunctionalization, or an increase in the rate of mutation. Distinguishing among these possibilities requires a case-by-case analysis—a proposition that is antithetic to the ethos of Big Science genomics and bioinformatics.

Because positive selection leaves a number of footprints on the genome, and each test is designed to pick up on a slightly different signal, researchers sometimes combine multiple metrics into composite tests toward the goal of providing greater power of detection and a finer spatial resolution. Scores of such tests are typically referred to as composite scores. Tests employing composite scores come in two distinct forms. First, some methods calculate a composite score for a contiguous genetic region rather than a single nucleotide site by combining individual scores at all the sites within the region. The motivation for such an approach is that, although false positives may occur at any one nucleotide site by chance, a contiguous region of positive markers is unlikely to be spurious and most likely represents a *bona fide* signature of selection (e.g., Carlson et al. 2005). In the other type of method, composite scores are calculated by combining the results of many tests at a single site. The purpose of these methods is to utilize complementary information from different tests in order to provide better spatial resolution and pinpoint selection to the root cause (e.g. Zeng et al. 2006; Grossman et al. 2010).

It is very important to note that regardless of method or combination of methods, there are factors that conspire to underestimate the functional fraction of the genome and factors that conspire to overestimate the functional fraction of the genome. That is, some of the genomic segments identified as functional through telltale signs of positive selections may be false positives, while others may elude detection. For example, functional sequences may be under selection regimes that are difficult to detect, such as positive selection or weak purifying selection. In addition, selection may be difficult to detect as far as recently evolved species-specific elements (e.g., Smith et al. 2004) or very short genetic elements are concerned (e.g., De Gobbi et al. 2006). These factors would cause the fraction of the genome that is under selection and, hence, functional to be underestimated. On the other hand, selective sweeps, background selection, and significant reductions in population size (bottlenecks) would cause an overestimation of the fraction of the genome that is under selection (e.g., Williamson et al. 2007).

*What Proportion of the Human Genome is Functional?*

While it is undisputed that many functional regions within genomes have evolved under complex selective regimes, such as selective sweeps, balancing selection, and recent positive selection, it is widely accepted that purifying selection persisting over long evolutionary times is the most common mode of evolution (Rands et al. 2014). Studies that identify functional sites by using the degree of conservation between sequences from two (or more) species have estimated the proportion of functional nucleotides in the human genome to be 3%-15% (Ponting and Hardison 2011; Ward and Kellis 2012). We note, however, that each lineage gains and loses functional elements over time, so the proportion of nucleotides under selection needs to be understood in the context of divergence between species. For example, estimates of constraint between any two species will only include sequences that were present in their common ancestor and that have not been lost, replaced, or nonfunctionalized in the lineages leading up to the genomes of the extant species under study. Functional element turnover is defined as the loss or gain of purifying selection at a particular locus of the genome, when changes in the physical or genetic environment, causes a locus to switch from being functional to being nonfunctional or vice versa.

By using genomic data from 12 mammalian species and an estimation model that takes into account functional element turnover, Rands et al. (2014) estimated that 8.2% of the human genome is functional, with a 95% confidence interval of 7.1-9.2%. Because of the difficulties in estimating the functional fraction of a genome, evolutionary biologists treat such numbers as somewhat underestimated. Thus, a claim that 10% or even 15% of the human genome is functional would, thus, be tolerable. On the other hand, a claim that 80% of the human genome is functional (e.g., ENCODE Project Consortium 2012) is misleading in the extreme and logically risible.

Unsurprisingly, in Rands et al.'s (2014) study, constrained coding sequences turned out to be much more evolutionary stable, i.e., experienced least functional element turnover, than constrained noncoding sequences. From among noncoding sequences, the sequences that were most likely to be functional were enhancers and DNase 1 hypersensitivity sites. Transcription factor binding sites, promoters, untranslated regions, and long noncoding RNAs (lncRNAs) contributed little to the functional fraction of the human genome, with lncRNAs exhibiting the most rapid rate of functional element turnover of all the noncoding element types. This finding implies that the vast majority of lncRNAs are devoid of function and represent transcriptional noise.

*How Much Garbage DNA Exists in the Human Genome?*

Because humans are diploid organisms and because natural selection is notoriously slow and inefficient in ridding populations of recessive deleterious alleles, the human genome is expected to contain garbage DNA. The amount of garbage DNA, however, should be quite small—many orders of magnitude smaller than the amount of junk DNA.

Deleterious alleles should exhibit a few telltale signs. First, they should be maintained in the population at very low frequencies. The reason for the rarity of deleterious alleles is that at low frequencies, the vast majority of such alleles will be found in heterozygous state, unexposed to purifying selection. Second, deleterious alleles should only very rarely become fixed between populations. Thus, one can identify them by using the ratio of polymorphic alleles to fixed alleles. Third, as shown by Maruyama (1974), deleterious or slightly deleterious allele should, on average, be younger than a neutral allele segregating at the same frequency. The young age of deleterious alleles is due to the fact that although purifying selection is not very efficient, it does eventually eliminate deleterious alleles from the population.

Many studies have shown that human genomes consist measurable quantities of garbage DNA. Tennessen et al. (2012), for instance, sequenced 15,585 protein-coding genes from 2,440 individuals of European and African ancestry, and found that out of the more than 500,000 single-nucleotide variants, the majority were rare (86% with a minor allele frequency less than 0.5%). Of the 13,595 single-nucleotide variants that each person carries on average, ~43% were missense, nonsense, affected splicing, i.e., affected protein sequence. The rest of the mutations were synonymous. About 47% of all variants (74% of nonsynonymous and 6% of synonymous) were predicted by at least one of several computational methods to be deleterious, and almost all of these deleterious variants (~97%) had very low frequencies. Fu et al. (2103) analyzed 15,336 protein-coding genes from 6,515 individuals, and estimated that ~73% of all single-nucleotide variants and ~86% of the variants predicted to be deleterious arose in the past 5,000-10,000 years. Sunyaev et al. (2001) estimated that the average human genotype carries approximately 1,000 deleterious nonsynonymous single nucleotide variants that together cause a substantial reduction in fitness.

Chun and Fay (2009) approached the problem of distinguishing between deleterious mutations from the massive number of nonfunctional variants that occur within a single genome by using a comparative genomics data set of 32 vertebrate species. They first identified protein-coding sites that were highly conserved. Next they identified amino acid variants in humans at protein sites that are evolutionarily conserved. These amino acids variants are likely to be deleterious. Application of this method to human genomes revealed close to 1,000 deleterious variants per individual, approximately 40% of which were estimated to be at allele frequencies smaller than 5%. Their method also indicated that only a small subset of deleterious mutations can be reliably identified.

So far, we have discussed population genetics and evolutionary methods for predicting the deleteriousness of protein-altering variants based on population properties. There are, however, methods that combine evolutionary and biochemical information to make such inferences (Cooper and Shendure 2011). Nonsense and frameshift mutations are the most obvious candidates, as they are predicted to result in a loss of protein function and are heavily enriched among disease-causal variation. However, this class of variation is not unambiguously deleterious, in some cases allowing functional protein production or resulting in a loss of protein that is apparently not harmful. Considering nonsynonymous variants, the simplest and earliest approaches to estimate deleteriousness was to use discrete biochemical categorizations such as "radical" versus "conservative" amino acid changes. However, there are now numerous more sophisticated approaches to classify nonsynonymous variants on both quantitative and discrete scales. These methods can be divided into "first-principles approaches" and "trained classifiers."

First-principles approaches explicitly define a biological property of deleterious variants and make predictions on the basis of similarity or dissimilarity to that property. For example, first principles approaches may use the presence of frameshifts in the coding regions as identifiers of deleteriousness (e.g., Sulem et al. 2015). By contrast, trained classifiers generate prediction rules by identifying heuristic combinations of many potentially relevant properties that optimally differentiate a set of true positives from a set of true negatives. First-principles approaches have the advantage of greater interpretability; for example, radical and conservative annotations of amino acid substitutions have a straightforward biochemical interpretation. However, first-principles methods are only as good as the assumptions that they make and do not model all of the relevant factors. Conversely, a trained classifier approach effectively yields a "black-box" prediction and will be prone to the biases and errors in the data. However, trained classifiers have the advantage of being specifically tunable to the desired task (such as, predicting disease causality) and are capable of incorporating many sources of information without requiring a detailed understanding of how that information is relevant.

We note that all methods for predicting deleteriousness are prone to estimation error. For example, all methods use multiple sequence alignments and phylogenetic reconstructions. Low qualities of alignment or erroneous phylogenetic tree may result in low-quality inferences. Moreover, the sampling of species is crucial. A sample consisting of sequences that are very similar to one another offers less power of detection, thus increasing the number of false negatives. Conversely, inclusion of distant sequences may increase the number of false positives.

Finally, we note that many methods exploit biochemical data, including amino acid properties (such as charge), sequence information (such as presence of a binding site) and structural information (such as the presence of a β-sheet). The integration of these data with comparative sequence analysis significantly improves predictions of deleteriousness.

*Mutational Origins of Junk DNA*

Here, we ask ourselves which among the numerous mutational processes can increase genome size and concomitantly increase the fraction of nonfunctional DNA in the genome. Increases in genome size can be caused by genome duplication, various types of subgenomic duplications, mononucleotide and oligonucleotide insertions, and replicative transposition. Genome size increases can occur either gradually or in a punctuated manner. There are no analogous processes that can cause large and sudden decreases in genome size. Thus, as opposed to genome-size increases, which can occur in fits and starts, decreases in genome size can only occur in a gradual manner.

Insertions and subgenomic duplications can increase genome size, however, these process are expected to increase the size of the genome only very slowly, such that their contribution to genome size is thought to be negligible. Moreover, none of these two processes is expected to alter the fraction of nonfunctional elements in the genome. For example, the probabilities of duplicating a functional sequence or a nonfunctional sequence are proportional to the prevalence of such sequences in the genome, so that following subgenomic duplication the fraction of junk DNA in the genome neither increases nor decreases.

Genome duplication is the fastest route to genome size increase; in one fell swoop, genome size is doubled. Genome duplication, however, does not increase the fraction of junk DNA in the genome unless the newly created functional redundancy is quickly obliterated by massive gene nonfunctionalization. Of course, in recently formed polyploids, one cannot speak of an increase in the C value, since this value refers to the size of the haploid genome and does not depend on the degree of ploidy. The contribution of gene duplication to the variation in C values only comes into play after the polyploid species has been diploidized and became a cryptopolyploid.

Because of their ability to increase in number, many transposable elements can have profound effects on the size and structure of genomes. Indeed, replicative transposable elements, especially retrotransposons, have the potential to increase their copy number while active, and many are responsible for huge genome size increases during relatively short periods of evolutionary time (e.g., Piegu et al. 2006).

The vast majority of eukaryotic genomes studied to date contain large numbers of transposable elements, and in many species, such as maize (*Zea mays*), the edible frog (*Pelophylax esculentus* formerly known as *Rana esculenta*), and the largest genome sequenced so far, the loblolly pine (*Pinus taeda*), transposable elements constitute the bulk of the genome (e.g., Kovach et al. 2010). Organisms devoid of transposable elements are extremely rare. The only group of organisms that is known to lack transposable elements altogether or to possess only 2-3 copies of transposable elements are yeast species, such as *Ashbyia gossipii*, *Kluyveromyces lactis*, *Zygosaccharomyces rouxii*, and *Schizosaccharomyces octosporus* (Dietrich et al. 2004; Rhind et al. 2011).

Replicative transposition is the only mutational process that can greatly and rapidly increase genome size, and at the same time significantly increase the fraction of junk DNA in the genome. The reason for this is that transcription and reverse transcription are very inaccurate methods of copying genetic information, and hence most of the increase in genome size due to transposable elements will consist of nonfunctional elements, i.e., junk DNA.

The claim that most junk DNA is made of transposable elements and their incapacitated descendants yields a quantitative prediction—it is expected that a positive correlation should exist between the genomic fraction inhabited by transposable elements and genome size. We note, however, that estimating the numbers and kinds of transposable elements in a genome is far from trivial or routine. First, different sequenced genomes differ from one another in the quality of the sequences. Because repetitive elements are difficult to sequence and problematic to use in genomic assemblies, the fraction of repetitive DNA is frequently underrepresented in low quality genome sequences. Second, available algorithms for detecting repeated elements are know to perform with varying degrees of success in different species. The reason is that algorithms use a database of known transposable sequences as their reference, so that the repertoire of transposable elements in one species may be better characterized that that of another species. Third, most transposable elements have neither been coopted into a function by the host nor retained their ability to transpose. Thus, they evolve in an unconstrained fashion, loosing their similarity to other members of their transposable-element family very rapidly. Algorithms that rely on similarity measures are, hence, unable to identify a significant proportion of dead transposable elements. Finally, most algorithms for detecting repeated transposable elements fail to discover short elements. As a consequence, the fraction of a genome that is taken up by transposable elements is more often than not extremely underestimated. For example, about 50% of the human genome has been identified as derived from transposable elements by using algorithms that rely on a database of consensus element sequences. By using a method that identifies oligonucleotides that are related in sequence space to one another, de Koning et al. (2011) found out that 66-69% of the human genome is derived from repetitive sequences.

Despite the difficulties described above, a positive correlation between total sequence length of transposable elements and genome size is seen in many groups. For example, the Pearson correlation coefficient in a sample of 66 vertebrate genomes was 0.77 (Tang 2015). In other words, ~60% of the variation in genome size could be explained by the variation in total length of transposable elements. In the literature, one can find many reports of positive correlations between the number of transposable-element copies per genome and genome size (Kidwell 2002; Biémont and Vieira 2006; Hawkins et al. 2006; Tenaillon et al. 2010; Lynch et al. 2011; Chénais et al. 2012; Chalopin et al. 2015). The rule is simple, large genomes have huge numbers of transposable elements; small genomes have very few (e.g., Roest Crollius et al. 2000; Kovach et al. 2010; Ibarra-Laclette et al. 2013; Kelley et al. 2014).

*Why so much of the genome is transcribed… or is it?*

The ENCODE Project Consortium (2012) consisted of approximately 500 researchers and cost in excess of 300 million dollars. Its purpose was to identify all functional elements in the human genome. One of its main findings was that 75% of the genome is transcribed—a phenomenon that was originally described by Comings (1972). Does this observation support the thesis that the human genome is almost entirely functional?

ENCODE systematically catalogued every transcribed piece of DNA as functional. In real life, whether a transcript has a function depends on many additional factors. For example, ENCODE ignored the fact that transcription is fundamentally a stochastic process that is inherently noisy (Raj and van Oudenaarden 2008). Some studies even indicate that 90% of the transcripts generated by RNA polymerase II may represent transcriptional noise (Struhl 2007). In fact, many transcripts generated by transcriptional noise may even associate with ribosomes and be translated (Wilson and Masel 2011). Moreover, ENCODE did not pay any attention to the number of transcripts produced by a DNA sequence. The vast majority of their newly "discovered" polyadenylated and non-polyadenylated RNAs are present at levels below one copy per cell and were found exclusively in the nucleus—never in the cytoplasm (Palazzo and Gregory 2014). According to the methodology of ENCODE, a DNA segment that produces 1,000 transcripts per cell per unit time is counted equivalently to a segment that produces a single RNA transcript once in a blue moon.

We note, moreover, that ENCODE used almost exclusively pluripotent stem cells and cancer cells, which are known as transcriptionally permissive environments. In these cells, the components of the RNA polymerase II enzyme complex can increase up to 1,000-fold, allowing for high transcription levels from random sequences. In other words, in these cells transcription of nonfunctional sequences, that is, DNA sequences that lack a *bona fide* promoter, occurs at high rates (Babushok et al. 2007). The use of HeLa cells is particularly suspect, as these cells have ceased long ago to be representative of human cells. For example, as opposed to humans who have a diploid number of 46, HeLa cells have a "hypertriploid" chromosome number, i.e., 76-80 regular-size chromosomes, of which 22-25 no longer resemble human chromosomes, in addition to an undetermined and highly variable number of "tiny" chromosomal fragments (Adey et al. 2013). Indeed,

HeLa has been recognized as an independent biological species called *Helacyton gartleri* (Van Valen and Maiorana 1991).

The human genome contains many classes of sequences that are known to be abundantly transcribed, but are typically devoid of function. Pseudogenes, for instance, have been shown to evolve very rapidly and are mostly subject to no functional constraint. Yet up to one-tenth of all known pseudogenes are transcribed (Pei et al. 2012) and some are even translated, chiefly in tumor cells (Kandouz et al. 2004). Pseudogene transcription is especially prevalent in pluripotent stem, testicular, germline, and cancer cells (Babushok et al. 2007). Unfortunately, because "functional genomics" is a recognized discipline within molecular biology, while "nonfunctional genomics" is only practiced by a handful of "genomic clochards" (Makalowski 2003), pseudogenes have always been looked upon with suspicion and wished away. Gene prediction algorithms, for instance, tend to "resurrect" pseudogenes *in silico* by annotating many of them as functional genes (Nelson 2004).

Another category of sequences that are devoid of function yet is transcribed are introns. When a human protein-coding gene is transcribed, its primary transcript contains not only functional reading frames but also introns and exonic sequences devoid of reading frames. In fact, only 4% of pre-mRNA sequences is devoted to the coding of proteins; the other 96% is mostly made of noncoding regions. Because introns are transcribed, ENCODE concluded that they are functional. But, are they? Some introns do indeed evolve slightly slower than pseudogenes, although this rate difference can be explained by a minute fraction of intronic sites involved in splicing and other functions. There is a long debate whether or not introns are indispensable components of eukaryotic genome. In one study (Parenteau et al. 2008), 96 introns from 87 yeast genes were removed. Only three of them (3%) seemed to have had a negative effect on growth. Thus, in the majority of cases, introns evolve neutrally, whereas a small fraction of introns are under selective constraint (Ponjavic et al. 2007). Of course, we recognize that some human introns harbor regulatory sequences (Tishkoff et al. 2006), as well as sequences that produce small RNA molecules (Hirose et al. 2003; Zhou et al. 2004). We note, however, that even those few introns under selection are not constrained over their entire length. Hare and Palumbi (2003) compared nine introns from three mammalian species (whale, seal, and human), and found that only about a quarter of their nucleotides exhibit telltale signs of functional constraint. A study of intron 2 of the human *BRCA1* gene, revealed that only 300 bp (3% of the length of the intron) is conserved (Wardrop et al. 2005). Thus, the practice of summing up all the lengths of all the introns and adding them to the pile marked "functional" is misleading.

The human genome is also populated by a very large number of transposable elements. Transposable elements, such as LINEs, SINEs, retroviruses, and DNA transposons, may, in fact, account for up to two-thirds of the human genome (de Koning et al. 2011) and for more than 31% of the transcriptome (Faulkner et al. 2009). Both human and mouse had been shown to transcribe *SINE*s (Oler et al. 2012). The phenomenon of *SINE* transcription is particularly evident in carcinoma cell lines, in which multiple copies of *Alu* sequences are detected in the transcriptome (Umylny et al. 2007). Moreover, retrotransposons can initiate transcription on both strands (Denoeud et al. 2007). These transcription initiation sites are subject to almost no evolutionary constraint, casting doubt on their "functionality." Thus, while a handful of transposons have been domesticated into functionality, one cannot assign a "universal utility" for all retrotransposons (Faulkner et al. 2009).

Whether transcribed or not, the majority of transposons in the human genome are merely parasites, parasites of parasites, and dead parasites, whose main "function" appears to be causing frameshifts in reading frames, disabling RNA-specifying sequences, and simply littering the genome.

*Hypotheses Concerning the Maintenance of Junk DNA*

Three main types of hypotheses have been proposed to explain the C-value paradox. In the first, the selectionist hypothesis, large eukaryotic genomes are assumed to be entirely or almost entirely composed of literal DNA. In the second hypothesis, the nucleotypic hypothesis, eukaryotic genomes are assumed to be almost entirely functional, but to contain mostly indifferent DNA and very little literal DNA. In neutralist hypotheses, genomes are assumed to contain large amounts of junk DNA—with bigger genomes containing larger amounts of junk DNA than smaller genomes.

*Selectionist Hypotheses*

Selectionist claims to the effect that genomes are entirely or almost entirely functional are usually made in the context of the human genome. For wholly unscientific reasons, humans are frequently assumed to occupy a privileged position, against which all other creatures are measured. In the literature, it frequently seems that as far as humans are concerned, the equations of population genetics are suspended and evolution abides by a different set of rules. Thus, while no one will ever insist that that ferns, salamanders, and lungfish, which have vastly larger genomes than humans, are devoid of junk DNA, one can frequently encounter National Institute of Health bureaucrats denying that human junk DNA exists (Zimmer 2015). It is in the human context, therefore, that we shall examine whether or not selectionist claims hold water.

The first selectionist hypothesis at the level of the genome was put forward by Zuckerkandl (1976), who asserted that the there is very little nonfunctional DNA in the genome; the vast majority of the genome performs essential functions, such as gene regulation, protection against mutations, maintenance of chromosome structure, and the binding of proteins. Consequently, the excess DNA in large genomes is only apparent. In the many years since the publication of this article, this theory was rejected multiple times and the paper was forgotten. The ENCODE Project Consortium (2012) resurrected the selectionist hypothesis (without acknowledging its originator), but their conclusion that the human genome is almost 100% functional was reached through various nefarious means, such as by employing the "causal role" definition of biological function and then applying it inconsistently to different biochemical properties, by committing a logical fallacy known as "affirming the consequent," by failing to appreciate the crucial difference between "junk DNA" and "garbage DNA," by using analytical methods that yield false positives and inflate estimates of functionality, by favoring statistical sensitivity over specificity, and by emphasizing statistical significance over the magnitude of the effect (Eddy 2012; Doolittle 2013; Graur et al. 2013; Niu and Jiang 2013; Brunet and Doolittle 2014; Doolittle et al. 2014; Palazzo and Gregory 2014).

Given that the original selectionist hypothesis has been appropriately relegated to the dustbin of history, it is difficult to rationalize the contemporary resurrection of such theories. Explaining the motives behind the misguided pronouncements of ENCODE Project Consortium (2012) would require the combined skills of specialists in the pathologies of ignorance, pseudoscience, and self-aggrandizement.

*Nucleotypic and Nucleoskeletal Hypotheses*

A variety of cellular, organismal, and ecological parameters have been reported to correlate with C values. Unfortunately, these relationships are never universal; they are apparent only in limited taxonomic contexts. For example, although genome size is inversely correlated with metabolic rate in both mammals and birds, no such relationship is found in amphibians. Many correlates of genome size were described in plants (Greilhuber and Leitch 2013), but have no applicability outside this kingdom. For example, species with large genomes flower early in the spring, while later flowering species have progressively smaller genomes. Additional examples concern weediness (the ability to invade arable lands) and invasiveness (the ability to colonize new environments), which were both found to be negatively correlated with genome size. An intriguing hypothesis for explaining genome size in plants casts phosphorus as a key player. Phosphorus is an important ingredient in the synthesis of nucleic acids (DNA and RNA). Yet despite its being the twelfth most abundant element in the environment, it is not readily accessible to plants and is often present in such low amounts that it may be considered a limiting nutrient for DNA synthesis. Because large genomes require increased supplies of phosphorus, it has been hypothesized that polyploids and plants with large genomes are at a selective disadvantage in phosphorus-limited environments (Hanson et al. 2003). Phosphorus-depleted soils should, accordingly, be populated by species with small genome sizes. Indeed, plants that live in mineral-poor environments seem to have particularly small genomes. Moreover, the smallest plant genome reported so far was found in a family of carnivorous plants that grow in nutrient-poor environments (Greilhuber et al. 2006). Some experimental support for the phosphorus hypothesis has been obtained by Šmarda et al. (2013) in their long-term fertilization experiment with 74 vascular plant species.

So far, the most universal correlate of genome size is cell size. For over a hundred years, cytologists have been aware of a positive correlation between the volume of the nucleus and the volume of the cytoplasm. These observations led to the concept of the nucleocytoplasmic ratio, according to which the ratio of the nuclear volume to that of the cytoplasm is constant, reflecting the need to balance nuclear and cytoplasmic processes. Indeed, neither nuclear-DNA content, nor varied growth conditions, nor drug treatments, could alter the nucleocytoplasmic ratio, and deviations from it are associated with disease (Jorgensen et al. 2007; Neumann and Nurse 2007).

Given that a positive correlation exists between genome size and nuclear volume and that cytoplasm volume and cell volume are similarly correlated, the nucleocytoplasmic ratio is frequently presented as a positive correlation between genome size and cell size. A rough correlation between C value and cell size has been noted in some of the earliest studies of genome size evolution (Mirsky and Ris 1951), and has since been confirmed in many groups of animals, plants, and protists. Indeed, a positive correlation with cell size is a most general feature of genome size, and the relationships between C value and cell size has been claimed to rank among the most fundamental rules of eukaryote cell biology (Cavalier-Smith 2005). We note, however, that the best correlations are found among closely related taxa. For example, while a significant correlation between pollen size and genome size was found in a sample of 16 wind-pollinated grass species, a large-scale analysis of 464 angiosperms failed to confirm the correlation (Greilhuber and Leitch 2013). These findings make it clear that genome size evolution cannot be understood without reference to the particular biology of the organisms under study. Finally, we note, that in many taxa, cell size and genome size do not seem to be correlated (e.g., Starostova et al. 2008, 2009).

Two basic explanations for the correlation between genome size and cell size have been put forward in the literature: the coincidence and the nucleotypic hypotheses. Under the coincidence hypothesis, most DNA is assumed to be junk and genome size is assumed to increase through mutation pressure. The increase in the amount of DNA in the genome is not, however, a boundless process. At a certain point, the genome becomes too large and too costly to maintain, and any further increases in genome size will be deleterious and selected against. In the coincidence hypothesis, it is assumed that bigger cells can tolerate the accumulation of more DNA (Pagel and Johnstone 1992). In other words, the correlation between genome size and cell size is purely coincidental. The cell and the nucleus are envisioned as finite containers to be filled with DNA.

Under the nucleotypic hypothesis, the genome is assumed to have a nucleotypic function, i.e., to affect the phenotype in a manner that is dependent on its length but independent of its sequence. As a consequence, genome size may be under secondary selection owing to its nucleotypic effects. Let us assume, for instance, that genome size affects flowering time. Then, the selection for earlier or later flowering times will result in an indirect selection on genome size.

Cavalier-Smith (1978; 1982; 1985; 2005) envisioned the DNA as a "nucleoskeleton" around which the nucleus is assembled, so that the total amount of DNA in a cell as well as its degree of packaging exerts a strong effect on nucleus size and subsequently on cell size. According to this nucleoskeletal hypothesis, the DNA is not only a carrier of genetic information, but also a structural material element—a nucleoskeleton that maintains the volume of the nucleus at a size proportional to the volume of the cytoplasm. Since larger cells require larger nuclei, selection for a particular cell volume will secondarily result in selection for a particular genome size. Thus, the correlation between genome size and cell size arises through a process of coevolution in which nuclear size is adjusted to match alterations in cell size.

In nucleotypic hypotheses, the entire genome is assumed to be functional, although only a small portion of it is assumed to be literal DNA. Thus, according to this hypothesis, most DNA is indifferent DNA, whose length is maintained by selection, while its nucleotide composition changes at random. With this hypothesis, The driving force, according to the nucleotypic hypothesis, is selection for an optimal cell size. For example, if a large cell size becomes adaptively favorable due to changes in the environment, then there would also be positive selection for a corresponding increase in nuclear volume, which in turn will be achieved primarily through either increases in the amount of indifferent DNA or modifications in the degree of DNA condensation.

At this point, we need to raise two questions: (1) does cell size matter? and (2) does genome size affect cell size? The answer to the first question is that cells do have an optimal size, i.e., they need to be not too big and not too small. Based on studies in *C. elegans* and other systems, it has been argued that cell size is limited by the physical properties of its components (Marshall et al. 2012). For example, in order to proliferate, a cell has to divide, and for faithful cell division, the molecular machinery, such as the centrosome (the organelle that serves as the main microtubule organizing center) and the mitotic spindle, must be constructed at the right position and with the correct size. This may not be accomplished in extremely large or extremely small cells due to the physical properties of microtubules and chromosomes. If a cell exceeds the upper size limit, its centrosome and mitotic spindle may not be able to position themselves at the center of the cell, leading to nonsymmetrical cell division. Moreover, in such a case, microtubules may not reach the cell cortex, potentially leading to insufficient spindle elongation. If a cell falls below a lower size limit, its centrosome may not be able to position itself in a stable manner at the center of the cell due to the excess elastic forces of the microtubules. In addition, there may not be sufficient space for accurate chromosome segregation.

We do not know at present what the upper or lower limits of cell size are. We do, however, know that some cells are so extremely large, that they most certainly exceed whatever the theoretical upper limit for cell size may be. Three such examples are the shelled amoeba, *Gromia sphaerica*, which can reach a diameter of 3 cm, ostrich eggs, which can reach 15 cm in diameter and can weigh more than 1 kg, and the record holders, unicellular seaweeds belonging to genus *Caulerpa*, whose tubular stolon may extend to a length of 3 or more meters. These enormous examples cannot, however, be taken as evidence that an upper limit for cell size does not exist. What these examples show is that there exist molecular and cellular devices for escaping the consequences of large cell size during cell division. None of these enormous cells undergoes regular binary fission; they either divide by cleaving only a small portion of their mass (e.g., bird eggs), or by becoming multinucleate for at least part of their life cycle and producing large numbers of diminutive progeny or gametes (e.g., *Gromia* and *Caulerpa*).

As to the second question—does genome size affect cell size?—the evidence is quite thin. First, the correlation between cell size and genome size is imperfect at best. Second, there is evidence for contributors to cell size other than DNA. Levy and Heald (2007) studied the regulation of nuclear size in two related frog species: *Xenopus laevis* and *X. tropicalis*. *X. laevis* is a larger animal than *X. tropicalis* and its cells are tetraploid. *X. tropicalis* is smaller and its cells are diploid. The two species also differ in another aspect: the cells and nuclei of *X. laevis* are larger. Because *Xenopus* nuclei can be assembled in a test tube using chromatin (DNA–protein complexes) and extracts of egg cytoplasm, one can test whether or not DNA has a role in determining cell size. Levy and Heald (2007) added sperm chromatin from either *X. laevis* or *X. tropicalis* to egg extracts from either *X. laevis* or *X. tropicalis*. They found that, although both extracts can trigger assembly of the nuclear envelope around the chromatin, the cytoplasmic extract from *X. laevis* forms larger nuclei than the *X. tropicalis* extract, regardless of the DNA used. This indicates that one or more cytoplasmic factors determine nuclear size while DNA content seems to have no discernable effect. Can we, therefore, state that the nucleoskeletal hypothesis has been invalidated? My answer is that it may be too early to discard the nucleoskeletal hypothesis, although one may certainly call into question its universality.

*The Neutralist Hypothesis*

As explained previously, a scientific hypothesis should spell out the conditions for its refutation. As far as the C-value paradox is concerned, the simplest scientific hypothesis is that the fraction of DNA that looks superfluous is indeed superfluous. The assumption that a vast fraction of the genome evolves in a neutral fashion means that this DNA does not tax the metabolic system of eukaryotes to any great extent, and that the cost (e.g., in energy, time, and nutrients) of maintaining and replicating large amounts of nonfunctional DNA is negligible.

The first researcher to suggest that part of the genome may lack function was Darlington (1937), who recognized the difficulties in getting rid of redundant DNA because of its linkage to functional genes: "It must be recognized that the shedding of redundant DNA within a chromosome is under one particularly severe restriction, a restriction imposed by its contiguity, its linkage with DNA whose information is anything but dispensable." The theme of redundancy was later adopted by Rees and Jones (1972) and by Ohno (1972), the latter being credited with popularizing the notion that much of the genome in eukaryotes consists of junk DNA. In particular, Ohno (1972) emphasized the interconnected themes of gene duplication and trial and error in genome evolution. Gene duplication can alleviate the constraints imposed by natural selection by allowing one copy to maintain its original function while the other accumulates mutations. Only very rarely will these mutations turn out to be beneficial. Most of the time, however, one copy will be degraded into a pseudogene. As Ohno (1972) put it, "The creation of every new gene must have been accompanied by many other redundant copies joining the ranks of silent DNA base sequences," and "Triumphs as well as failures of nature's past experiments appear to be contained in our genome." The discovery of transposable elements, and more importantly the observation that vast majority of transposable elements are nonfunctional and highly degraded by mutation, added support for the junk DNA hypothesis, as did the discovery of other nonfunctional genomic elements, such as pseudogenes, introns, and highly repetitive DNA.

Under the neutralist hypothesis, we need to address three issues. First, we need to ask what mutational processes can create junk DNA, i.e., what does junk DNA consist of? Second, we need to identify the evolutionary driving forces that can maintain junk DNA. Third, we need to explain why different organisms possess vastly different quantities of junk DNA. The first question is relatively easy to answer. Notwithstanding the fact that gene and genome duplications can create redundancies that may ultimately result in junk DNA, and that satellite DNA can too contribute to the nonfunctional DNA fraction, there is currently no doubt whatsoever that the bulk of junk DNA is derived from transposable elements. By estimating the relative contribution of the major types of transposable elements, genomes can be classified into four main categories: (1) genomes in which with DNA transposons predominate (e.g., *Amphioxus*, *Ciona*, most teleost fish, *Xenopus*), (2) genomes in which *LINE*s and *SINE*s predominate (e.g., lamprey, elephant shark, *Takifugu*, coelacanth, chicken, mammals), (3) genomes with a predominance of LTR retrotransposons (e.g., the tunicate *Oikopleura*), and (4) genomes in which no particular transposable-element type predominates (e.g., *Tetraodon*, stickleback, reptiles, zebra finch). Some genomes are particularly poor in DNA transposable elements and contain almost exclusively retroelements (e.g., elephant shark, coelacanth, birds, mammals), however, there are no genomes that contain almost exclusively DNA transposable elements (Chalopin et al. 2015).

*Selfish DNA*

The "selfish-DNA" hypothesis attempts to explain how superfluous and even deleterious elements can multiply within genomes and spread within populations. Selfish DNA (Doolittle and Sapienza 1980; Orgel and Crick 1980) is a term that applies to DNA sequences that have two distinct properties: (1) the DNA sequences form additional copies of themselves within the genome, and (2) the DNA sequences either do not make any specific contribution to the fitness of the host organism or are actually detrimental. Some selfish DNA also engage in transmission ratio distortion and horizontal gene transfer—two processes through which they can further increase their frequency in the population.

The vast majority of selfish DNA comprises of class-I and class-II transposable elements that are (1) active, i.e., have the ability to produce copies of themselves by replicative transposition, and (2) have not been domesticated, i.e., coopted into function. A minor fraction of selfish DNA consists of promiscuous DNA as well tandemly repeated sequences. As an approximation, in the following, we shall use the terms "selfish DNA," "selfish DNA elements," and "transposable elements" interchangeably.

Because of their ability to increase in number, selfish DNA elements can have profound effects on the size and structure of genomes. Two main classes of hypotheses have been put forward to explain the long-term persistence of selfish DNA in the genome. One class of hypotheses proposes that the process reflects two independent equilibria. At the genome level, a balance is achieved between transposition or retroposition, on the one hand, and the mechanisms utilized by the host to restrain transposable-element activity, on the other. At the population level, a balance is achieved between the rate with which new copies of transposable elements are created and the efficiency with which negative selection gets rid of genotypes carrying transposable elements. The efficiency of selection, in turn, depends on population genetic parameters such as effective population size (Brookfield and Badge 1997; Le Rouzic et al. 2007; Levin and Moran 2011). Under this class of hypotheses, a genome is assumed to be a closed system, in which the activity of transposable elements is counteracted by such intrinsic entities as small interfering RNAs, PIWI-interacting RNAs (piRNAs), DNA methylation, and histone modifications. In this closed system, transposable elements employ various evasive strategies, such as preferential insertion into regions transcribed by RNA polymerase II. Sooner or later, the system comprised of the host genome and the transposable elements reaches a stable equilibrium, after which nothing much happens until internal or external perturbations, such as mutations or environmental stress, disrupt the equilibrium, at which point either a burst of transposition is unleashed or the transposable elements become forever incapacitated.

The discovery that some transposable elements, notably the *P* element of *Drosophila*, are able to colonize new genomes by means of horizontal transfer (Daniels et al. 1990) unveiled an additional way for transposable elements to persist over evolutionary time. Horizontal escape of an active transposable element into a new genome would allow the element to evade a seemingly inevitable extinction in its original host lineage resulting from host elimination or inactivation due to mutational decay (Hartl et al. 1997).

Although the inherent ability of transposable elements to integrate into the genome suggested a proclivity for horizontal transfer (Kidwell 1992), the extent to which such processes affected a broad range of transposable elements and their hosts was not clear. Schaack et al. (2010) revealed more than 200 cases of horizontal transfer involving all known types of transposable elements, which may mean that virtually all transposable elements can horizontally transfer.

From a genome-wide study across *Drosophila* species, it was estimated that approximately one transfer event per transposable-element family occurs every 20 million years (Bartolome et al. 2009). In several instances, we have evidence for horizontal transfer of transposable elements among extremely distant taxa, including at least 12 movements across phyla. So far, all these "long jumps" were found to involve DNA transposons, suggesting one of two possibilities: either DNA elements are better adapted to invade genomes than RNA elements, or the preponderance of DNA elements represents a case of ascertainment bias due to the fact that DNA elements are studied more intensively in an evolutionary context, while the research on RNA elements is almost exclusively focused on a narrow taxonomic range of so-called model organisms.

It is becoming increasingly clear that the lifecycle of a transposable-element family is akin to a birth-and-death process in that it starts when an active copy colonizes a novel host genome and it ends when all copies of the transposable-element family are lost or inactivated by chance through the accumulation of disabling mutations or by negative selection in a process which may be driven by host-defense mechanisms or by the fact that each transposable element contributes negatively to the fitness of the organism.

There are two major ways for transposable elements to escape extinction: the first is to horizontally transfer to a new host genome prior to extinction; the second is to inflict minimal fitness harm. Like other parasites, it is possible that transposable elements will make use of different strategies at different times. Each strategy has a phylogenetic signature. In cases in which horizontal transfer is frequent, there should be dramatic incongruence between the phylogeny of the transposable element family and that of their various host species. In these cases, horizontal transfer might allow the transposable element to colonize a new genome in which host suppression mechanisms are inefficient.

In cases where a transposable-element family has persisted for long periods in a host lineage, the reduced frequency of horizontal transfer can be inferred from the similarity between the phylogenies of the transposable element and the hosts. Persistence could be achieved, for instance, through self-regulatory mechanisms that limit copy number or by evolving targeting preference for insertion into "safe havens" in the genome, such as for instance through preferential transposition into high copy-number genes or heterochromatin. The *LINE-1* element of mammals provides an exceptional example of endurance, having persisted and diversified over the past 100 million years with virtually no evidence of horizontal transfer.

*The Mutational Hazard Hypothesis: A Nearly Neutralist Hypothesis*

So far, we have provided plausible explanations for the persistence of junk DNA within genomes. We did not, however, address the question of genome size disparity: Why do certain organisms possess minute quantities of junk DNA, whereas the genome of other eukaryotic organisms is made almost entirely of nonfunctional DNA. A possible explanation may be that some genomes have not been invaded by transposable elements while others have been invaded many times. Another explanations may be that some genomes have been invaded by very inefficient transposable elements while others by very prolific ones. A third explanation may be that some genomes are extremely efficient at warding off selfish DNA while others are more permissive. Since none of these three explanations yield predictions concerning the relationship between the amount of junk DNA and population genetic variables, to the best of our knowledge they have not been tested. The mutational-hazard theory (Lynch and Conery 2003; Lynch 2006) uses differences in selection intensity and selection efficacy to explain the observed difference between organisms with high junk-DNA content and those with low junk-DNA content.

The mutational-hazard theory (Lynch and Conery 2003; Lynch 2006) postulates that virtually all increases in genome size in eukaryotes impose a slight fitness reduction. Thus, eukaryotic genomes are assumed to contain varying amounts of excess DNA that behaves like junk DNA in species with large effective population sizes and like garbage DNA in species with small effective population sizes.

The fate of a slightly deleterious allele is determined by the interplay between purifying selection and random genetic drift. Purifying selection acts by decreasing the frequency of the slightly deleterious allele at a rate that depends upon its selective disadvantage, $s$, where $s < 0$. Random genetic drift changes the allele frequencies in a non-directional manner at a mean rate that is proportional to $1/N_e$, where $N_e$ is the effective population size. Thus, whether a mutation that increases genome size is selected against or has the same probability of becoming fixed in the population as a neutral mutation is determined by its selective disadvantage relative to the effective size of the population into which it is introduced. Lynch and Conery (2003) argued that the ineffectiveness of selection in species with low $N_e$ is the key to understanding genome-size evolution, and the main prediction of the mutational-hazard theory is that large genomes will be found in species with small effective population sizes. How can we test this hypothesis?

First, we must ascertain that two mutational requirements are met. The first requirement is that mutations resulting in genome increases will outnumber mutations resulting in genome diminution. In many prokaryotes, for instance, this condition is not met. As a consequence, random genetic drift causes the genomes of many prokaryotes with small effective population sizes to get smaller rather than larger. In most eukaryotes, the proliferation of transposable elements overwhelms all other mutations and, hence, at the mutation level, the condition that genome increases outnumber genome decreases is met.

The second mutational requirement is that the addition of each transposable element to the genome should on average decrease the fitness of its carrier. Two types of empirical data support this assumption. First, it has been shown that in *Drosophila melanogaster*, each transposable element insertion decreases the fitness of an individual by 0.4% (Pasyukova et al. 2004). Second, it has been shown that bottlenecks affect the diversity and level of activity of transposable elements, with populations that had experienced population size reductions having an increased level of both transposable-element activity and transposable-element diversity (Lockton et al. 2008; Picot et al. 2008).

After determining that the mutational requirements of the mutational-hazard hypothesis are met, we need to ascertain that the postulated relationship between genome size and effective population size ($N_e$) is supported by empirical data. In principle, $N_e$ can be estimated directly by monitoring the variance of allele-frequency changes across generations, as this has an expected value $\frac{p(1-p)}{2N_e}$, where $p$ is the initial allele frequency. In practice, however, as the expected changes in allele frequency from generation to generation are extremely small unless $N_e$ is tiny, this approach is difficult to put into practice because errors in estimating $p$ will overwhelm the true change in $p$ unless the sample size is enormous. As a consequence, most attempts to estimate $N_e$ have taken a circuitous route, the most popular being to indirectly infer effective population size from the levels of within-population variation at nucleotide sites assumed to evolve in a neutral fashion. The logic underlying this approach is that if $\mu$ is the rate of neutral point mutations per generation per site and if $N_e$ is roughly constant, then an equilibrium level of variation will be reached in which the mutational input to variation, $2\mu$, is matched by the mutational loss via genetic drift, $1/(2N_e)$. At equilibrium, the nucleotide site diversity is expected to be approximately equal to the ratio of these two forces, i.e., $4N_e\mu$ for a diploid population and $2N_e\mu$ for haploids. Thus, to estimate effective population size, one needs to empirically determine the degree of nucleotide-site diversity, which is a relatively straightforward process, and the rate of mutation, $\mu$, which is a slightly more difficult thing to do.

By measuring nucleotide site variation at synonymous sites and taking into account the contribution from mutation, average $N_e$ estimates turned out to be ~$10^5$ for vertebrates, ~$10^6$ for invertebrates and land plants, ~$10^7$ for unicellular eukaryotes, and more than $10^8$ for free living prokaryotes (Lynch et al. 2011). Although crude, these estimates imply that the power of random genetic drift is substantially elevated in eukaryotes—e.g., at least three orders of magnitude in large multicellular species relative to prokaryotes. It is also clear that the genetic effective sizes of populations are generally far below actual population sizes.

At present there is insufficient data on effective population sizes from a sufficiently diverse sample of taxa to test the mutational-hazard hypothesis directly. There is, however, indirect evidence supporting it. The mutation rate per generation is expected to be higher in species with low effective population sizes than in species with large effective population sizes. The mutation-hazard hypothesis asserts that organisms with low effective population sizes should have larger genomes than organisms with large effective population sizes. Thus, support for the mutational-hazard hypothesis can be obtained indirectly by showing that a positive correlation exists between mutation rate and genome size. Such a correlation has indeed been reported by Lynch (2010).

*Genome size and bottlenecks: The simultaneous accumulation of* Alu*s, pseudogenes, and* numt*s within primate genomes*

As far as slightly deleterious mutations are concerned, the importance of random genetic drift relative to selection becomes especially pronounced during profound reductions of population size. If genome size can be shown to have increased concomitantly with reductions in population size, then the mutational-hazard theory would gain evidential support.

Gherman et al. (2007) compared the evolution of *numt* insertions in primate genomes, and compared it with the insertions of two other classes of nonfunctional elements, *Alu*s and pseudogenes (Britten 1994; Bailey et al. 2003; Ohshima et al. 2003). These elements are unlikely to be functional, since their rate of evolution indicates a complete lack of selective constraint and they possess no positional, transcriptional, or translational features that might indicate a beneficial function subsequent to their integration into the nuclear genome. Using sequence analysis and fossil dating, Gherman et al. (2007) showed that a probable burst of integration of *Alu*s, pseudogenes, and *numt*s in the primate lineage occurred close to the prosimian–anthropoid split, which coincided with a major climatic event called the Paleocene–Eocene Thermal Maximum (~56 million years ago). During this event, which lasted for about 70,000 years, average global temperatures increased by approximately 6°C, massive amounts of carbon were added the atmosphere, the climate became much wetter, sea levels rose, and many taxa went extinct or suffered decreases in population size. Thus, the increase in primate genomes can be largely accounted for by a population bottleneck and the subsequent neutral fixation of slightly deleterious nonfunctional insertions. The fact that three classes of nonfunctional elements that use vastly different mechanisms of multiplying in the genome increased their numbers simultaneously, effectively rules out selectionist explanations. These findings suggest that human and primate genomic architecture, with its abundance of repetitive elements, arose primarily by evolutionary happenstance.

*Is it junk DNA or is it indifferent DNA?*

Distinguishing between neutralist and nucleoskeletal explanations has been quite difficult. Pagel and Johnstone (1992) proposed two expectations derived from each of the two theories. According to these authors, a major cost of junk DNA is the time required to replicate it. Organisms that develop at a slower pace may therefore be able to "tolerate" greater amounts of junk DNA, and thus a negative correlation across species between genome size and developmental rate is predicted. In contrast, the prediction of the nucleoskeletal hypothesis is for a positive correlation between genome size and cell size. Unfortunately, organisms with large cells also tend to develop slowly, whereas faster-growing organisms typically have smaller cells. Thus, according to the skeletal DNA hypothesis, a negative correlation between developmental rate and the C value is also expected. However, according to the nucleotypic hypothesis, the relation between developmental rate and genome size occurs only secondarily, as a result of the relationship between developmental rate and cell size.

Pagel and Johnstone (1992) studied 24 salamander species. The size of the nuclear genome was found to be negatively correlated with developmental rate, even after the effects of nuclear and cytoplasmic volume have been removed. However, the correlations between genome size, on the one hand, and nuclear and cytoplasmic volumes, on the other, become statistically insignificant once the effects of developmental rates have been removed. These results support the hypothesis that most of the DNA in salamanders is junk rather than indifferent DNA. Whether Pagel and Johnstone's results represent a true phenomenon or one restricted to *Salamandra* is still a controversial subject (Gregory 2003), especially since in eukaryotes, the "cost" of replicating DNA may not be correlated with genome size.

TABLE 1. C values of a few eukaryotic organisms ranked by genome size.

| Species[1] | C value (Mb) |
|---|---|
| **Saccharomyces cerevisiae (baker's yeast)** | **13** |
| *Caenorhabditis elegans* (nematode) | 78 |
| *Ascidia atra* (sea squirt) | 160 |
| *Drosophila melanogaster* (fruit fly) | 180 |
| **Paramecium aurelia (ciliate)** | **190** |
| *Oryza sativa* (rice) | 590 |
| *Strongylocentrotus purpuratus* (sea urchin) | 870 |
| **Gymnosporangium confusum (rust fungus)** | **893** |
| *Gallus domesticus* (chicken) | 1,200 |
| *Lampetra planeri* (brook lamprey) | 1,900 |
| *Boa constrictor* (snake) | 2,100 |
| *Canis familiaris* (dog) | 2,900 |
| **Homo sapiens (human)** | **3,300** |
| *Nicotiana tabacum* (tobacco plant) | 3,800 |
| *Locusta migratoria* (migratory locust) | 6,600 |
| **Paramecium caudatum (ciliate)** | **8,600** |
| *Schistocerca gregaria* (desert locust) | 9,300 |
| *Allium cepa* (onion) | 15,000 |
| **Coscinodiscus asteromphalus (centric diatom)** | **25,000** |
| *Lilium formosanum* (lily) | 36,000 |
| *Psilotum nudum* (skeleton fork fern) | 71,000 |
| *Amphiuma means* (two-toed salamander) | 84,000 |
| *Pinus resinosa* (Canadian red pine) | 68,000 |
| *Protopterus aethiopicus* (marbled lungfish) | 130,000 |
| *Paris japonica* (canopy plant) | 150,000 |

[1] Unicellular organisms and humans are listed in bold letters.

Source: Graur (2016) and references therein.

Legends to Figures

Figure 1. A classification of genomic segments by biochemical activity. Each of the three categories can be functional or functionless (rubbish).

Figure 2. An evolutionary classification of genomic elements according to their selected-effect function. From Graur et al. (2015).

Figure 3. A nomenclature for some possible changes in the functional affiliation of genomic elements. From Graur et al. (2015).

Figure 1.

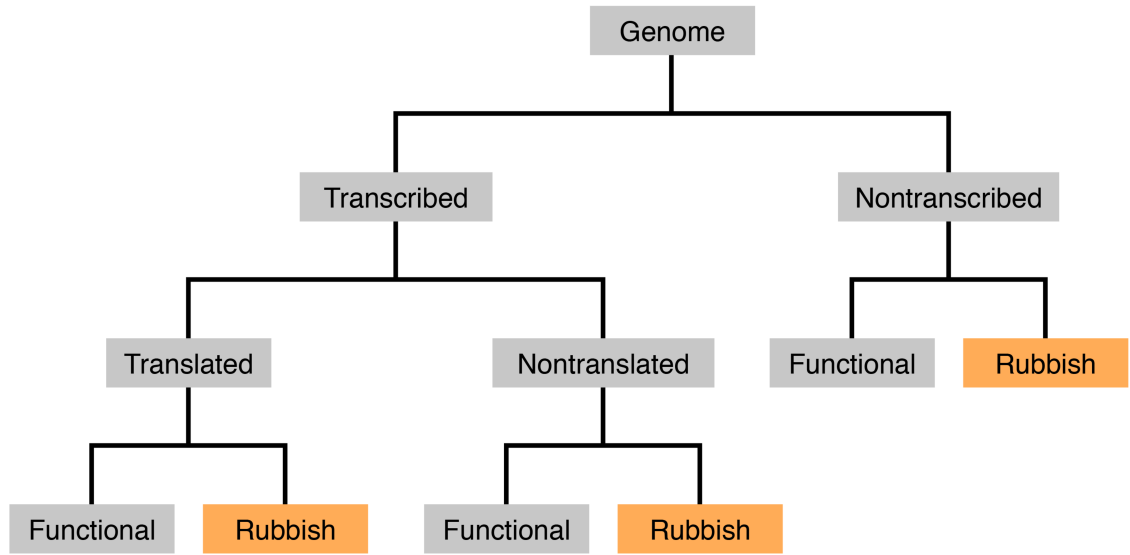

Figure 2.

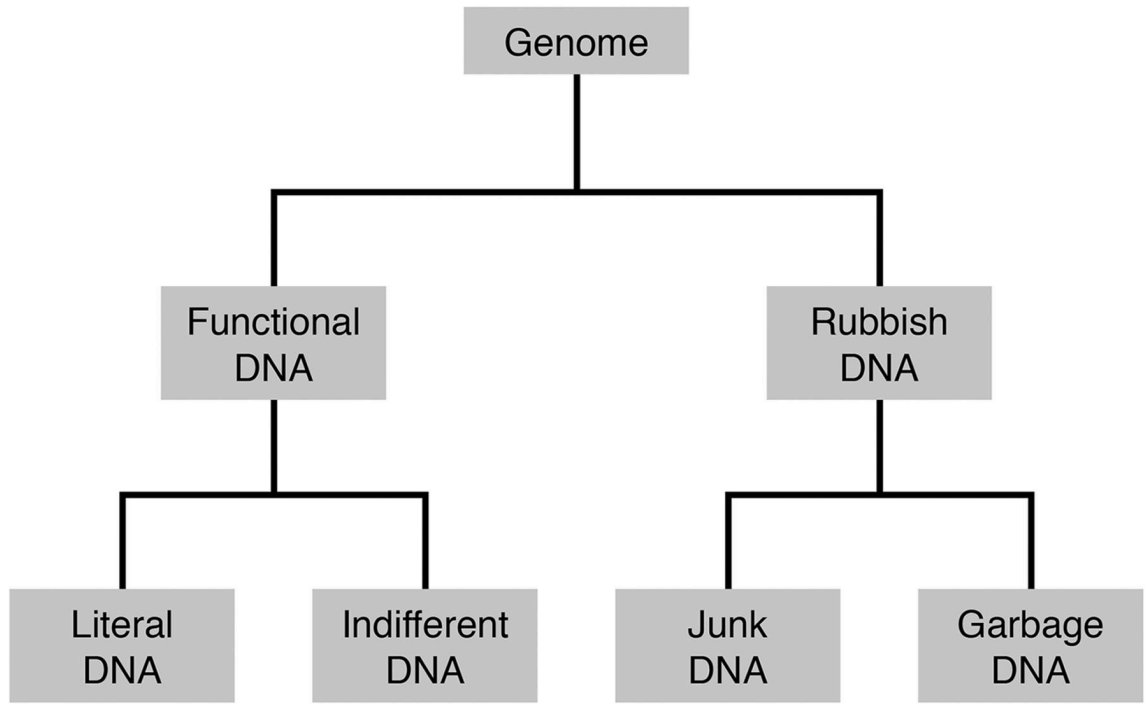

Figure 3.

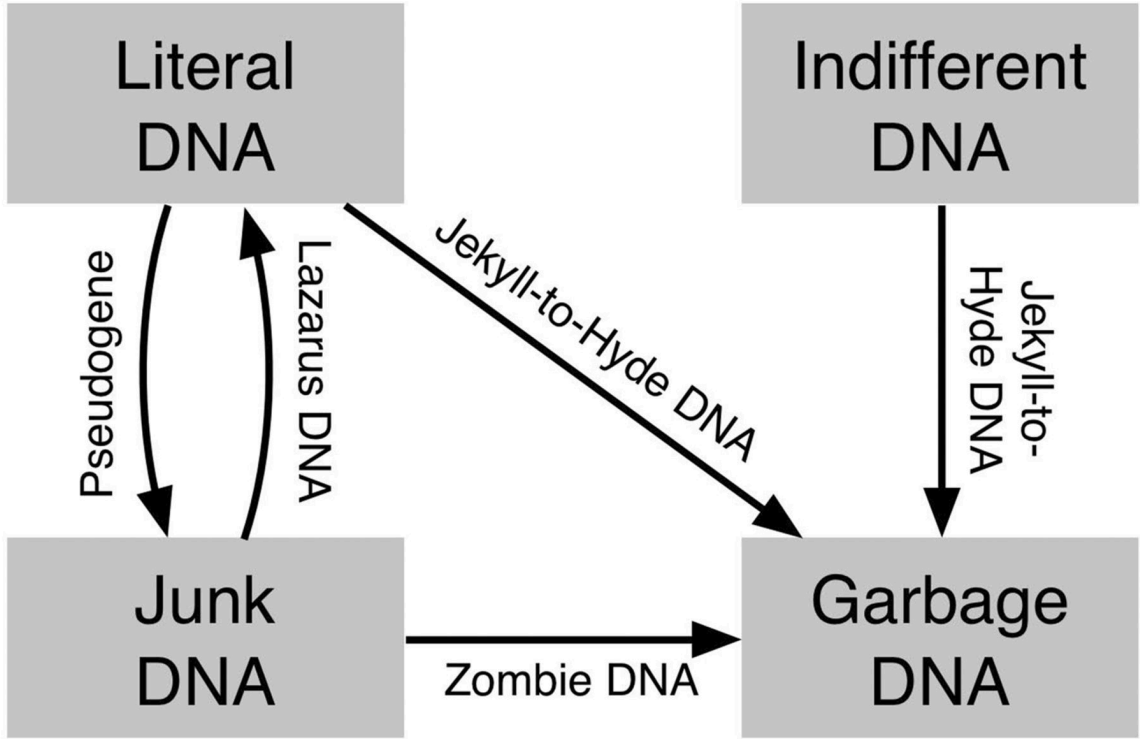